\documentclass[epj]{svjour}
\usepackage{graphicx}
\usepackage{hyperref}
\hypersetup{colorlinks = true, allcolors = blue}
\usepackage[utf8]{inputenc}
\usepackage{authblk}
\usepackage{hepparticles}
\usepackage{hepunits}
\usepackage{hepnames}
\usepackage{amsmath}
\usepackage{amssymb}
\usepackage{textgreek}
\begin{document}

\title{Challenges for FCC-ee Luminosity Monitor Design}

\author{Mogens Dam
}                     
\offprints{}          
\institute{Niels Bohr Institute, Copenhagen University, Denmark}
\date{Received: \today / Revised version: \today}
%
\abstract{
For cross section measurements, an accurate knowledge of the integrated luminosity is required. The FCC-ee Z lineshape programme sets the ambitious precision goal of $10^{-4}$ on the \emph{absolute} luminosity measurement and one order of magnitude better on the \emph{relative} measurement between energy-scan points.
The luminosity is determined from the rate of small-angle Bhabha scattering, $\mathrm{e^+e^- \to e^+e^-}$, where the final state electrons and positrons are detected in dedicated monitors covering small angles from the outgoing beam directions.
The constraints on the luminosity monitors are multiple: 
\mbox{\emph{i}) they} are placed inside the main detector volume only about 1\,m from the interaction point;
\mbox{\emph{ii})} they are centred around the outgoing beam lines and do not satisfy the normal axial detector symmetry;
\mbox{\emph{iii})} their coverage is limited by the beam pipe, on the one hand, and the requirement to stay clear of the main detector acceptance, on the other;
\mbox{\emph{iv})} the steep angular dependence of the Bhabha scattering process imposes a geometrical precision on the acceptance limits at about 1\,\textmugreek rad, corresponding to geometrical precisions of 
$\mathcal{O}(1\,\text{\textmugreek m})$; and
\mbox{\emph{v})} the very high bunch crossing rate of 50\,MHz during the Z-pole operation calls for fast readout electronics.
Inspired by second-generation LEP luminosity monitors, a proposed ultra-compact solution is based on a sandwich of tungsten-silicon layers. A vigorous R\&D programme is needed in order to ensure that such a solution satisfies the more challenging FCC-ee requirements.
\PACS{
      {PACS-key}{describing text of that key}   \and
      {PACS-key}{describing text of that key}
     } 
} 
\maketitle
\section{Introduction}
\label{intro}

The integrated luminosity acts as the link between the number of produced events, $N$, and the cross section, $\sigma$, for any physics process:
\begin{equation}
    N = L \cdot \sigma .
    \label{eq:N=Lsigma}
\end{equation}
Precise cross-section measurements therefore depend on the accurate knowledge of the luminosity.
In hadron-hadron collisions,
the main method for luminosity measurement is via direct calculations based on known beam parameters, with important input on transverse beam profiles from so-called \emph{van der Meer} scans.
Precisions at the few percent level are reached at the LHC using this method~\cite{GRAFSTROM201597}.
In electron-positron collisions, the luminosity is usually inferred by exploiting Eq.\ (\ref{eq:N=Lsigma}) for a reference process with a precisely know cross section, $\sigma_\mathrm{ref}$. This way, cross-section measurements are, in fact, based on the ratio of event counts between the physic process under study and the reference process,
$\sigma = (N/N_\mathrm{ref}) \cdot \sigma_\mathrm{ref}$. In order for measurements
not to be limited by the statistical uncertainty on $N_\mathrm{ref}$, the reference process should be chosen with a sufficiently large cross section, $\sigma_\mathrm{ref} \gtrsim \sigma$.
At the recent generation of e$^+$e$^-$ colliders, the B-factories operating at  center-of-mass energies around the $\Upsilon$(4S) resonance ($\sim$\,10 GeV), 
sub-percent level precisions have been reached based on large angle production of e$^+$e$^-$, \textgamma\textgamma{}, or \textmu$^+$\textmu$^-$ final states~\cite{Lees:2013rw,Abudinen:2019osb}.

Luminosity measurements with precisions at the sub-per-mille level were pioneered 
at LEP using small-angle Bhabha scattering,  e$^+$e$^- \rightarrow$ e$^+$e$^-$.
Events were detected by dedicated monitors encircling the beam at about 2.5 m from the interaction point. The small-angle Bhabha
cross section can be calculated with high precision from quantum electrodynamics and depends only weakly on the properties of the Z boson, even at center-of-mass energies close to the Z pole.
To lowest order, the strongly forward-peaked cross section takes the form
\begin{equation}
    \sigma = \frac{16\pi\alpha^2}{s}
    \left(
    \frac{1}{\theta_\mathrm{min}^2} - \frac{1}{\theta_\mathrm{max}^2}
    \right),
\end{equation}
for a detector with a polar angle coverage from $\theta_\mathrm{min}$ to $\theta_\mathrm{max}$ and full coverage in azimuth. 
During the first period of LEP operation, it was realised that the \mbox{Z $\rightarrow$ hadrons} process could be measured with very small systematic uncertainty well below the per-mille level. An upgrade of the luminosity monitors, originally designed for percent level measurements, therefore became desirable.
Upgrades took different form in the four experiments: In L3, a precise 
Si tracker was added in front of their BGO calorimeter~\cite{Brock:1996ty}; In DELPHI, the original SPACAL-like calorimeter was replaced by a more precise device based on scintillating tiles~\cite{Camporesi:1997uxj}; In OPAL and ALEPH, new
compact \mbox{Si-W} sandwich calorimeters were added in front of the original devices covering the lowest scattering angles~\cite{Opal.Lumi,Aleph.Lumi}. 
To appreciate the extraordinary efforts that were invested in reaching sub-per-mille levels of accuracy, it is worthwhile consulting the work by OPAL, described in detail in Ref.\ \cite{Opal.Lumi}, resulting in an experimental precision as low as 
$3.4 \times 10^{-4}$, a factor two or more better than the LEP competitors.
%
%
At this stage, dominant contributions to the remaining uncertainty included: 
\emph{i}) radial metrology of the calorimeters
\mbox{($1.4 \times 10^{-4}$)}, 
\emph{ii}) uncertainty on the correspondence between measured shower coordinates and true scattering angles \mbox{($1.4\times 10^{-4}$)},
\emph{iii}) calorimeter energy response \mbox{($1.8\times 10^{-4}$)}, and \mbox{\emph{iv}) clustering} algorithm \mbox{($1.0\times 10^{-4}$)}. Smaller contributions were associated with 
\mbox{\emph{v}) beam related backgrounds} \mbox{($0.8\times 10^{-4}$)} and
\mbox{\emph{vi}) beam} parameters 
\mbox{($0.6\times 10^{-4}$)}.

At FCC-ee, precise luminosity measurement will again be of vital importance. The FCC-ee programme 
includes four major phases with precision measurements of the four heaviest particles of the Standard Model: 
\emph{i})~the Z boson from $5\times 10^{12}$ Z decays collected around the Z pole, 
\emph{ii})~the W boson from $10^8$ WW pairs collected close to threshold, 
\mbox{\emph{iii})}~the Higgs boson from $1.2\times 10^6$ \mbox{e$^+$e$^- \to$ HZ} events produced at the  cross-section maximum, and 
\emph{iv})~the top quark from $10^6$ $\mathrm{t\bar{t}}$ pairs produced at and slightly above threshold. 
In particular, the two first phases, with their superior  statistics, call for the highest achievable systematic accuracy. Ambitious precision goals have been set at 10$^{-4}$ for the 
%
absolute luminosity measurement and one order of magnitude better for the relative measurement between energy scan points.
%
%
As at LEP, the luminosity measurement is foreseen to be principally based on small-angle Bhabha scattering. 
This may be complemented by large-angle $\mathrm{e^+e^-} \to${} \textgamma\textgamma{} production, where, despite a three orders of magnitude smaller rate, the statistical precision is sufficient for most purposes, and the systematic uncertainties will be entirely different. 

This essay concentrates on the luminosity measurement based on small-angle Bhabha scattering; the design of the luminosity monitors and the measurement methodology. Emphasis is placed on deriving the required geometrical precisions of the monitors and of their alignment with respect to the accelerator in order to facilitate the absolute luminosity precision goal. The design described is that developed for the FCC-ee Conceptual Design Report~\cite{Benedikt:2651299}.

\section{Experimental environment}

\begin{figure}
\begin{center}
\includegraphics[width=0.52\textwidth]{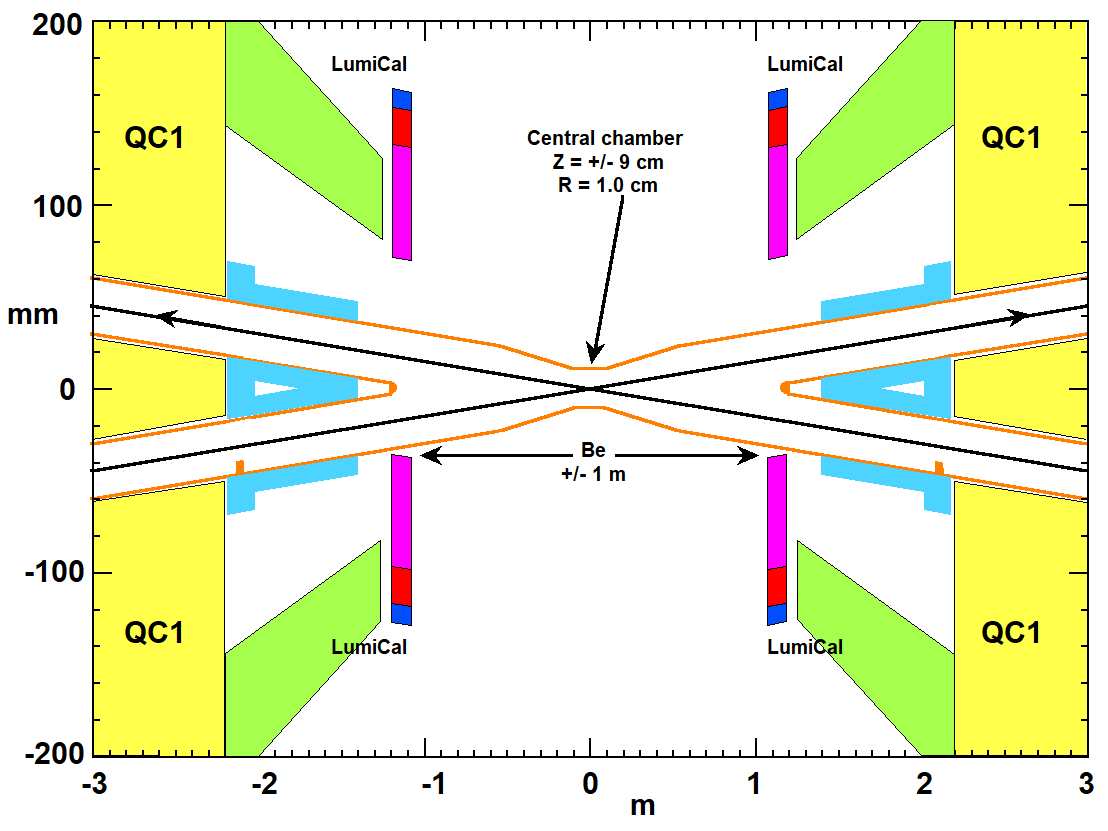}
\caption{Sketch of the interaction region in the horizontal plane.
Compensating solenoids (green) are installed in front of the quadupoles (yellow) to compensate for the traversal of the beams at an 15 mrad angle through the detector field. 
The luminosity calorimeters (LumiCal) are installed in front of the compensating solenoids and centered around the outgoing beam lines. 
\mbox{From \cite{Boscolo:2021hsq}}.}
\label{fig:ir}
\end{center}
\end{figure}

The experimental environment at FCC-ee is described in detail in the CDR~\cite{Benedikt:2651299}.
The FCC-ee accelerator has been designed to provide optimal luminosities at all energies from the Z pole to the \mbox{$t\bar{t}$ production} threshold.
Of direct relevance to the luminosity measurement, the design includes: \emph{i})~separate storage rings for the electron and positron beams, which are brought into collision at a 30 mrad horizontal crossing angle at two (possibly four) interaction points (IP); \emph{ii}) strong vertical focusing of the beams provided by a set of quadrupoles the closest with its face at \mbox{$L^* = 2.2$ m} from the IP; \emph{iii}) very high beam crossing rates of $\mathcal{O}$(50~MHz) at \mbox{Z-pole} operation. A detector is placed at each IP, with a solenoid that delivers a magnetic field of 2~T parallel to the bisector of the two beams, the $z$-axis. Two complementary detector concepts have been studied. In both cases, the trajectories of charged particles are measured within a tracker down to polar angles of about 150 mrad with respect to the $z$ axis. The tracker is surrounded by a calorimeter and muon-detection system. The region covering polar angles below 100 mrad corresponds to the ``machine-detector interface" (MDI), the design of which demands special care. 
The crossing of the beam lines at angles of \mbox{$\pm$15 mrad} with
the detector field necessitates the insertion of a set of compensating solenoidal magnets in front of the quadrupoles in order to preserve the low vertical beam emittance. This arrangement pushes the luminosity monitors far into the main detector volume, where the available space is severely constrained. A brief account of the MDI can be found in Ref.\ \cite{Boscolo:2021hsq}, from which \mbox{Fig.\ \ref{fig:ir}}, showing a sketch of the interaction region, is taken.

%

\section{Luminosity Monitor Design}

A set of Si-W calorimeters is proposed as luminosity monitors at FCC-ee. In addition to providing a very compact solution, the choice of a purely calorimetric measurement also has theoretical advantages.
As it has been pointed out by the authors of the higher-order Bhabha event generator \textsc{Bhlumi}~\cite{BHLUMI}, there is theoretical difficulty in ascribing meaning to the trajectory of a bare Bhabha-scattered electron. Theoretically more meaningful is the concept of a dressed electron, i.e.\ an electron including its close-lying radiated photons. This favours a calorimetric measurement. 

\begin{figure}
\begin{center}
\includegraphics[width=0.6\textwidth]{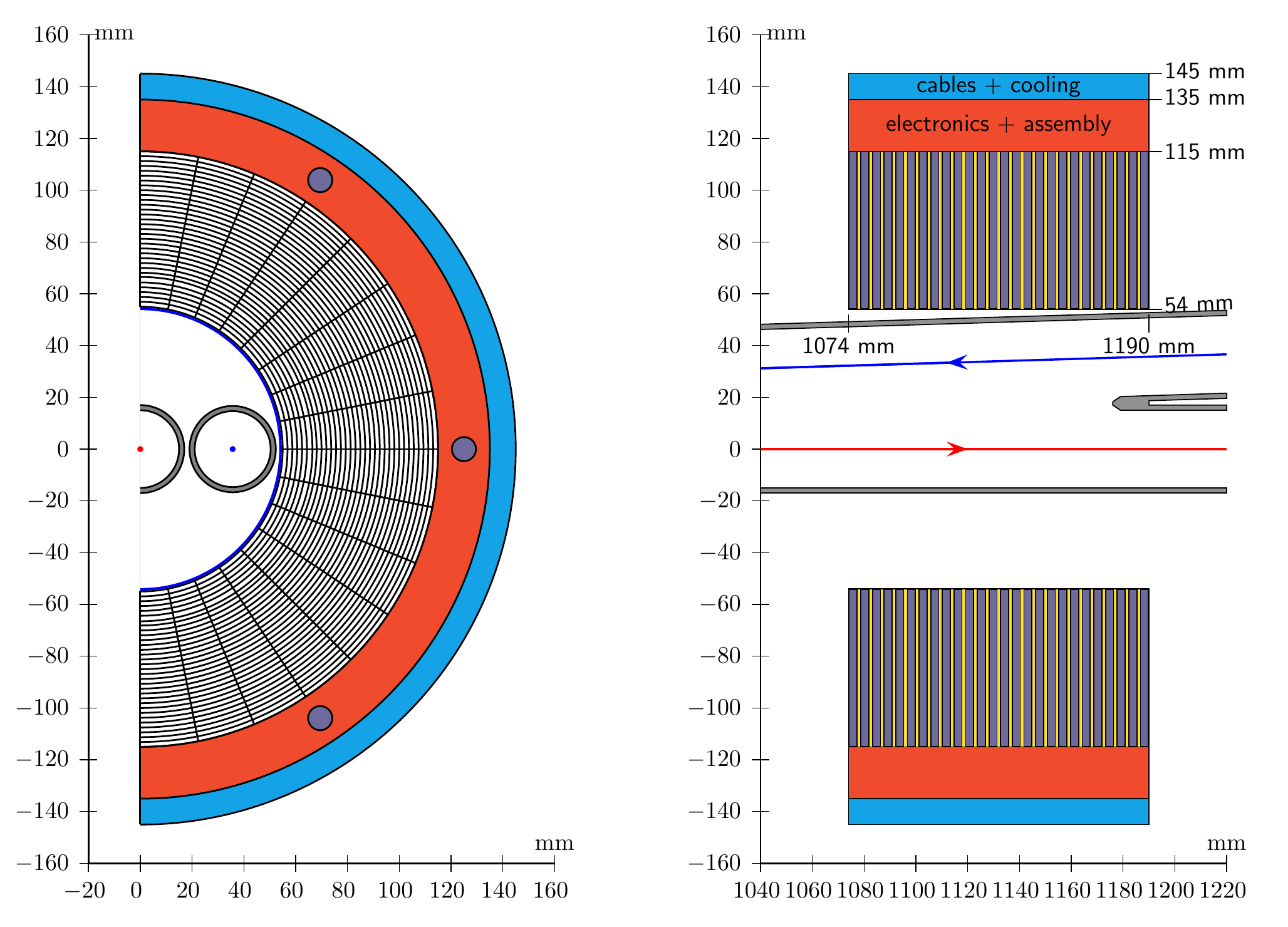}
\caption{The luminosity calorimeter centred around the outgoing beam line (shown in red): front view (left), top view (right).}
\label{fig:luminometer}
\end{center}
\end{figure}

Following the examples from OPAL~\cite{Opal.Lumi} and ALEPH~\cite{Aleph.Lumi} and from linear collider studies~\cite{ilc_det_tdr,Aicheler:2012bya}, the proposed 
calorimeters have been designed as cylindrical devices assembled from stacks of identical Si-W layers. This simple geometry facilitates the control of construction and metrology tolerances to the necessary micron level. 
To measure precisely the scattering angles of the Bhabha-scattered electrons and positrons, the calorimeters are centred around---and tilted to be perpendicular to---the outgoing beam lines.
The calorimeters sit in front of the compensating solenoids extending to $|z| \simeq 1.2$~m from the IP.
Their physical dimensions are then limited by two parameters: At their inner radius, they must stay clear of the incoming beam pipe; at their outer radius, they must stay away from the tracker acceptance defined by a $150$-mrad cone around the $z$ axis. The situation is illustrated in \mbox{Fig.\ \ref{fig:luminometer}}, where also the main physical dimensions of the proposed design are given. The design includes 25 layers, each layer comprising a 3.5-mm tungsten plate, equivalent to $1\,X_0$, and a Si-sensor plane inserted in the $1$-mm gap. In the transverse plane, the Si sensors are finely partitioned into pads. The proposed number of divisions is 32 both radially and azimuthally for 1\,024 readout channels per layer, or 25\,600 channels in total for each calorimeter. A 30-mm uninstrumented region at the outer circumference is reserved for services. This includes front-end electronics, cables and cooling, and,
importantly, also the physical structures---likely including precision dowels and bolts---for the assembly of the Si-W sandwich. It should be emphasised that a proper engineering design has so far not been done but is highly needed in order to verify the proposed assembly procedure and the space necessary for this at the outer radius.
Overall, the proposed design is very compact, each calorimeter weighing only about $65$~kg.

In the base design, each calorimeter is divided vertically into two half barrels clamped together around the beam pipe. Due to the small dimensions, it will be possible to produce each silicon half-layer from a single silicon tile, which will minimise potential inactive regions between sensors and facilitate the precise geometrical control of the acceptance. Meticulous care is required for the design of the vertical assembly of the two half-barrels, both in order to avoid a non-instrumented region and to precisely control the geometry. An alternate design has been proposed, where the monitors would be built as full barrels avoiding the vertical assembly. 
This would have strong implications for the design of the MDI and the assembly procedure of the detector around the beam pipe.

The Si-sensor pads will be connected to compact front-end electronics positioned at radii immediately outside the sensors. To minimise the occurrence of pile-up events---at the Z pole, the average number of Bhabha events hitting the luminometers per bunch crossing is about 10$^{-3}$---it is desirable to read out the detector at the bunch-crossing rate. This calls for the development of readout electronics with a shaping time below the bunch spacing of $20$~ns. A power budget of $5$~mW per readout channel has been estimated, for a total of $130$~W per calorimeter, to be removed via cooling. With thermal expansion coefficients of both W and Si-crystal at the 
\mbox{few\,$\times 10^{-6}/\,^\circ$C} level, control of the monitors' temperature within a $1\, ^\circ$C tolerance would seem appropriate for control of physical dimensions to the 
\mbox{1-\textmu{}m} level. Measurements from OPAL~\cite{Opal.Lumi}, however, show a considerably higher sensitivity of about 4 \textmu{}m\,/\,$^\circ$C for their radial dimensions. This difference has to be understood. As in OPAL, temperatures should be continously monitored and gradients should be minimised.

\section{Acceptance and Geometrical Precisions}
\label{sec:lumimeast}


The proposed luminometer design has a full-depth coverage for scattering angles between 50 and 97 mrad. For a robust energy measurement, the acceptance limits must be kept some distance away from these borders. Choosing this distance to be 15~mm, corresponding to one Moli\`ere radius,  limits the acceptance to the 62--88~mrad range. To ensure that the luminosity measurement has no first order dependence on possible misalignments and movements of the beam system relative to the luminometer system (both in position and angle), the method of asymmetric acceptance, first introduced in \mbox{Ref.\ \cite{CRAWFORD1975173}} and later extensively used at LEP, will be employed. Bhabha events are selected if the ${\rm e^\pm}$ is inside a narrow acceptance in one calorimeter, and the ${\rm e^\mp}$ is inside a wide acceptance in the other. A 2~mrad difference between the wide and narrow acceptances is judged adequate to accommodate possible misalignments and movements of the IP. The narrow acceptance thus covers the 64--86~mrad angular range, corresponding to a Bhabha cross section of 14~nb at the Z pole (compared to 40~nb for Z production), equivalent to $6.4 \times 10^{-4}$ events per bunch crossing.

The forward-peaked $1/\theta^3$ spectrum of the Bhabha process causes the luminosity measurement to be particularly sensitive to the determination of the angular coverage. The sensitivity can be determined by simple analytic calculations, using only the $1/\theta^3$ spectrum and the
physical dimensions of the monitor system.
The Bhabha acceptance $A$ is affected by any change $\Delta R_{\rm in}$ ($\Delta R_{\rm out}$) of the inner (outer) edge radial coordinate as follows:
\begin{equation}
    \frac{\Delta A}{A} \approx 
    -\frac{\Delta R_\mathrm{in}}{1.6\,\text{\textmu{}m}} \times 10^{-4},
    \qquad \text{and} \qquad 
    \frac{\Delta A}{A} \approx 
    +\frac{\Delta R_\mathrm{out}}{3.8\,\text{\textmu{}m}} \times 10^{-4}. 
\end{equation}
Similarly, $A$ is affected by any change $\Delta Z$ of the half-distance between the effective planes defining the radial measurements in the two calorimeters:
\begin{equation}
\frac{\Delta A}{A} \approx +\frac{\Delta Z}{55\,\text{\textmu{}m}} \times 10^{-4}. 
\end{equation}
With the 30~mrad beam crossing angle, the situation becomes slightly more complicated.
Now, with the two calorimeters centred on different axes, $Z$ shall be interpreted as
$Z = \frac{1}{2}(Z_1+Z_2)$, where $Z_1$ and $Z_2$ are the distances, measured along the two outgoing beam axes, from the (nominal) IP to the two luminometers. 

With the method of asymmetric acceptance, only second-order dependencies of $A$ on the average IP position remains. Again the strengths of these dependencies can be estimated analytically. This gives a rather precise estimate for transverse offsets. However, longitudinal offsets imply an implicit cut into the acollinearity distribution of Bhabha events and thus a dependence on the amount of initial state radiation. To quantify this, a high-statistics study was performed based on the Bhabha event generator \textsc{Bhlumi}~\cite{BHLUMI} and a parameterised detector response, where the energy of close-lying electrons and photons were collected into clusters. The study confirmed the second order dependencies as long as offsets were sufficiently small to be covered by the difference between the wide and narrow acceptance definitions: in this case, offsets of up to $\delta r \approx 0.5$~mm transversely and $\delta z \approx 20$~mm longitudinally. Inside this range, the observed changes of the acceptance could be parameterised as 
\begin{equation}
\frac{\Delta A}{A} \approx 
+\left(\frac{\delta r}{0.6\,\text{mm}}\right)^2 \times 10^{-4} 
\qquad \text{and} \qquad 
\frac{\Delta A}{A} \approx 
-\left(\frac{\delta z}{6\,\text{mm}}\right)^2 \times 10^{-4}.
\label{eq:offsets}
\end{equation}
Whereas the result for transverse offsets agrees well with the analytic estimate, it is interesting to notice that, for longitudinal offsets, the simulation result is larger (by a factor four) than that of the analytic estimate and even has the opposite sign, demonstrating the importance of radiative effects. The effect of a possible tilt of the luminometer system with respect to the FCC-ee beam-line system has a similar effect on $A$ as a transverse offset: In Eq.\ \ref{eq:offsets} one simply has to replace $\delta r$ with $Z \delta \alpha$, where $\alpha$ is the tilt angle.
It should be stressed that such offsets and tilts give rise to asymmetries in the Bhabha counting rate either azimuthally (radial offsets and tilts) or between the two calorimeters (longitudinal offsets) and can therefore be monitored and corrected for directly from the data. No such possibility of correction from the data exists for the detector construction tolerances, $\Delta R$ and $\Delta Z$, discussed in the previous paragraph, which therefore need to be controlled via precise metrology and alignment.
As a reference, OPAL~\cite{Opal.Lumi} achieved control of the inner radius of their calorimeters to a precision of 4.4 \textmu{}m through precise metrology. Relative to OPAL, several uncertainty contributions are expected to vanish if each sensor layer can be made from a single silicon crystal. A dominant remaining contribution then stems from the stability of the half-barrel separation, which was 1.9 \textmu{}m in the case of OPAL.

In summary, to reach a precison of $10^{-4}$ on the absolute luminosity measurement, the radial dimensions of the luminosity monitors have to be controlled to a precision of $\mathcal{O}(1\ \text{\textmu{}m})$, 
whereas the half-distance between the two monitors has to be controlled to
$\mathcal{O}$(50 \textmu{}m).
The requirements on the alignment of the luminometer system with respect to the average IP position are considerably more relaxed: accuracies of order 0.5~mm and 5~mm are called for in the radial and longitudinal directions, respectively. 

\section{Other sources of systematics}


A complete study of systematic effect is still to be pursued. Here we report on initial studies of effects associated with the accelerator: backgrounds and beam-beam effects. Evidently, both of these studies shall be followed up when a better understanding the accelerator behaviour becomes available.

At LEP, the primary source of background for the luminosity measurement was from coincidences of off-momentum particles generated by beam particles scattering with the residual gas in the beam pipe 
and deflected by the quadrupoles into the luminometers~\cite{Opal.Lumi}. With its much stronger focusing, the ratio between the instantaneous luminosity and the beam current is much higher at FCC-ee than at LEP. It is therefore not surprising, as pointed out already in 
\mbox{Ref.\ \cite{Benedikt:2651299}}, that this background source appears negligible in FCC-ee. In the same reference, other sources of backgrounds such as those from incoherent pair production and from synchrotron radiation were likewise argued to be unimportant. 

Electromagnetic effects caused by the very large charge densities of the FCC-ee beam bunches affect the colliding particles in several related ways. The final state electrons and positrons from small angle Bhabha scattering are focused by the electromagnetic fields of the opposing bunches leading to a sizeable bias of the luminometer acceptance that must be corrected for. This effect was studied in \mbox{Ref.\ \cite{Voutsinas:2019eyu}}, where it was demonstrated that several sets of measurements can be used to control this bias, and that it therefore should not compromise the targeted precision. Interestingly, as recently pointed out in \mbox{Ref.\ \cite{Voutsinas:2019hwu}}, this effect was already present at LEP, where it led to an underestimation of the luminosity measurement by about 0.1\%, significantly larger than the originally reported experimental uncertainties.

\section{Relative Luminosity}

For the relative normalisation between energy scan points, geometrical effects, the main focus of this essay, tend to cancel. In OPAL, dominant effects on the relative normalisation were related to the accelerator with contributions from beam related backgrounds \mbox{($0.8\times 10^{-4}$)} and beam parameters \mbox{($0.6\times 10^{-4}$)}. Even if it is believed that FCC-ee compared to LEP will have less beam induced background and that the beam parameters (beam tilts and divergences) will have less freedom to vary over time (as an illustration the beam-pipe radius was $~$50 mm at LEP versus 15 (possibly 10) mm at FCC-ee) it will certainly be a formidable challenge to improve on LEP by one order of magnitude. As always, statistics is our friend facilitating precise control studies.

\section{Outlook}

Much work is needed towards a consolidated luminometer design fulfilling the many severe requirements. For the cylindrical CDR design, discussed in this essay, important points include:
\begin{enumerate}
    \item Engineering-level study of the proposed detector assembly method involving precision dowels and through-going bolts. This shall take into account the required $\mathcal{O}$(1 \textmu{}m) geometrical precision on the radial coordinate.
    \item Realistic estimate of the space needed for services at radii outside the sensitive region. This region shall be kept as transparent as possible by the use of light materials to minimise particle interactions.
    \item Design of a procedure for maintaining and monitoring the geometrical accuracy of the monitors via precise metrology and alignment. 
    \item Design of compact, low-power readout electronics preferentially allowing readout at the 50 MHz bunch-crossing rate. This shall include the transmission of signals away from the detectors.
    \item Design of a cooling system allowing control of a constant and uniform temperature over the monitors. The required tolerances have to be developed.
    \item Integration of the monitors into the MDI design including support structures, which isolate the monitors from possible movements and vibration of the accelerator magnet system;
\end{enumerate}

More fundamental issues are related to the detector coverage, which is severely limited by the geometrical constraints arising from the placement inside the main detector volume in the very crowded forward region.
\begin{enumerate} \addtocounter{enumi}{6}
\item
There is no coverage for scattering angles below 50~mrad. This goes against the wish to close as hermetically as possible FCC-ee detectors down towards the beam line. Certainly there is physical room at the inside of the CDR design to place instrumentation towards smaller scattering angles in all azimuthal directions except that of the incoming beam pipe. The question is how such irregular shaped instrumentation would be compatible with the 1-\textmu{}m level accuracy needed on the radial limits of the acceptance definition. A engineering-level study is needed to answer this. 
\item There is full-depth coverage only up to scattering angles of 96 mrad. In the main detector system, this corresponds to a polar angle of 81 mrad in the azimuthal direction of the incoming beam.
This is considerably lower than the agreed-upon boundary at \mbox{100 mrad} between the MDI and the detector. Hence, there is a risk that the forward electromagnetic calorimeter, with its face at $|z| \simeq 2.3$\,m, will not be able to reach down to such small angles and provide the necessary overlap between the two systems. At the same time, the outermost ``corner" of the CDR design, with the \mbox{30 mm} reserved for services, reaches out to a polar angle of \mbox{150 mrad}. Should it be necessary to extend the service region further, this risks to clash with the tracking system. There is a clear tension here, and the final design will have to be done taking the overall detector layout into account.
\end{enumerate}


%
%
%

%
\bibliographystyle{myutphys}
\bibliography{references}

\end{document}